\documentclass[10pt,letterpaper]{article}
\usepackage{opex3}

\begin{document}

\title{Control of dispersion in photonic crystal waveguides using group symmetry theory}

\author{Pierre Colman,Sylvain Combri\'e,Ga\"{e}lle Lehoucq,Alfredo De Rossi}
\address{Thales Research and Technology, Avenue A. Fresnel \\ 91767 Palaiseau, France}

\email{alfredo.derossi@thalesgroup.com}



\begin{abstract}
We demonstrate dispersion tailoring by coupling the even and the odd modes in a line-defect photonic crystal waveguide. Coupling is determined ab-initio using group theory analysis, rather than by trial and error optimisation of the design parameters. A family of dispersion curves is generated by controlling a single geometrical parameter. This concept is demonstrated experimentally on a 1.5mm-long waveguide with very good agreement with theory. \end{abstract}



The linear and nonlinear propagation of waves depends crucially on the dispersion. The ability to control the dispersion is therefore a highly desired property of the physical medium supporting the propagation. For instance, the minimization of chromatic distortion in optical fibers \cite{agrawalNLoptics} is the main concern in optical communications. Moreover, some nonlinear interactions e.g. parametric conversion/generation, involving different frequencies, are completely determined by the dispersion, through the phase matching condition\cite{Boyd_NLO}.
\newline
\indent In optical fibers and waveguides, dispersion control is possible because the waveguide dispersion, which can be controlled through the geometry, can compensate for the material dispersion, which is material-dependent \cite{Yin:06}. In an optical waveguide, e.g. the so-called photonic wires, the dispersion can be controlled by changing the height and the width of the high index silicon wire on top of the silica layer \cite{Yin:06}. The ability to generate a prescribed dispersion function, also referred as \textit{dispersion engineering} \cite{Li:08,Benisty:09}, is a more demanding goal. By offering many degrees of freedom in the design, photonic crystals (PhC) fibers\cite{KnightNat2003} and waveguides \cite{Mori:04,Petrov:05,frandsen:06,Li:08,Hamachi:09} have proven to enable dispersion engineering. 
\newline 
\indent The particular dispersion of line-defect photonic crystal waveguides is often described as the result of an
anti-crossing between index-guided (e.g. guided by total internal reflection) and gap-guided modes (which only exist in PhCs) \cite{Petrov:04}. However the inherent complexity of photonic crystals, as opposed to simpler optical structures, limits the understanding of the mechanisms underlying the control of the dispersion. As a result, in order to tune a PhC waveguide, the design is optimized regardless of the nature of the underlying mechanism, e.g. the changes in the mode field distribution or in the coupling between modes. This said, dispersion engineering has been demonstrated in PhC waveguides by optimizing the coupling of two waveguides\cite{Petrov:05,Mori:04} or by coupling modes with different polarizations \cite{patterson_2009b}.
\newline
\indent In this paper we discuss a design procedure that, rather than optimizing some of the parameters defining the PhC waveguide, defines a suitable perturbation which optimizes the anti-crossing of two, well-defined, modes. Especially breaking the natural symmetry of the waveguide in order to couple Odd and Even mode appears as an effective way to control the dispersion in PhCs \cite{Lu:10,Mao:09,Ma:08}; still the symmetry should not be broken on a trial and error procedure. We demonstrate here that to obtain the desired dispersion features, depending on the initial design, the symmetry should be broken in a very specific way. This is illustrated by experimental results.\\ As an example, we consider here the even and odd modes, relative to the plane parallel to the propagation direction and perpendicular to the PhC slab. Those two modes are typically found in line-defect PhC waveguides and are uncoupled because they have different symmetries; however, the odd mode is rather regarded as a nuisance limiting the spectral extent of single mode operation. Here, we show that, owing to even-odd mode coupling, a variety of useful dispersion profiles can be controlled by a single parameter, which is a quite desirable feature.


To fix the ideas, let us consider first the case of two coupled PhC waveguides\cite{Mori:04} within the formalism of coupled mode theory (CMT). The dispersion of the super-mode resulting from the coupling $\gamma(k)$ of the modes of the individual waveguides, with dispersion $\omega_{wg\{1,2\}}(k)$, results from the diagonalisation of:  
\begin{equation}
\left| {\begin{array}{cc}
 \omega_{wg\{1\}}(k) & \gamma(k)  \\
 {\gamma}^*(k) & \omega_{wg\{2\}}(k)  \\
 \end{array} } \right| 
\end{equation}
Introducing the coupling results into a modification of the dispersion of the original modes, which can be controlled within some extent by playing with some parameters, such as the radius of the holes or the separation between the two waveguides. We point out that $\gamma(k)$ could be modified, in principle, almost independently from $\omega_{wg\{i\}}~i=\{1,2\}$.
\newline
\indent 
Here, we explore the possibility of coupling the two modes of a single waveguide in order to control the dispersion. Operating with a single waveguide, rather than a set of coupled waveguides, has many practical advantages: the excitation (i.e. the coupling) of the mode is easier and the linear and nonlinear effective areas are smaller, resulting into stronger light-matter interaction, a desirable feature in nanophotonics. 

\begin{figure}[htbp]
\includegraphics[width=12cm]{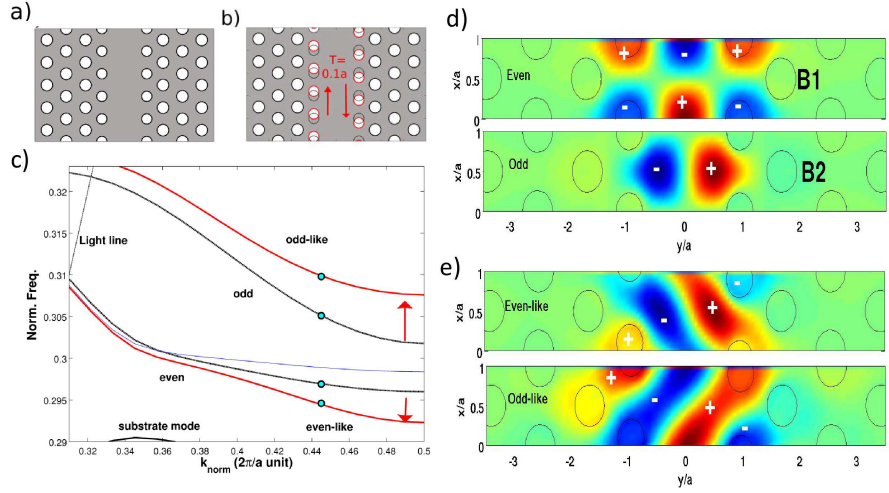}
\caption{PhC waveguide without (a) and with (b) perturbation. c) Corresponding band diagram of the initial (perturbed) structure in black (red) line. The cyan markers indicate the modes for which the field distribution is shown d)-e). The thin blue line corresponds to the dispersion obtained for a different perturbation as discussed in Fig. \ref{GammaCalcul}. d-e) Spatial distribution of the $H_z$-field, calculated at \textbf{K}=0.445, d) Even mode and Odd mode. e) even-like and odd-like mode.}
\label{Dispersion}
\end{figure}

Specifically, we consider the odd mode, defined relative to the plane of symmetry $x-z$, where $x$ is the direction of the propagation and $z$ is perpendicular to the 2D PhC lattice. 
\newline
\indent
Let us consider a line-defect waveguide with dispersion represented in fig. \ref{Dispersion}c, calculated using the 3D plane-wave expansion method implemented in the MPB code\cite{MPB}. The corresponding parameters, normalized with the lattice period, are: radius $r/a=0.26$, slab thickness $h/a=0.41$, width of the line defect $W/a=0.95\surd3$, outward displacement of the first row of holes $s/a=0.14$, with radius reduced to $r_1/a=0.23$. The refractive index of the material is assumed to be $3.17$. Here, the parameters have been chosen in order to lower the edge of the odd mode (at
$k=0.5\mathbf{K}$, \textbf{K} being the reciprocal lattice vector); indeed, here the frequency spacing is only $\delta f\,a/c=6\,$x$10^{-3}$. If a suitable coupling between these two modes, which are now closer in frequency, is introduced, then the anti-crossing will result into a change in the dispersion. As a consequence, controlling the coupling strength would enable, for example, the flattening of the lower frequency branch.


\indent Proceeding in the same way as for the coupled PhC waveguides, we now consider the coupling of even and the odd mode within the formalism of the coupled mode theory. The related eigenvalue problem is formulated by including the \textit{substrate} mode which is mainly of the index-guided type. This is necessary because that mode might also interact with the other modes. The off-diagonal terms represent the coupling induced by a suitable modification of the PhC design. Namely the purpose here is to get a simple way to control the dispersion by tuning only the interaction between the odd and the even modes, without suffering from possible interactions with the other modes. 
\begin{equation}
\left| {\begin{array}{ccc}
 \omega_{odd}(k) & \gamma_{oe}(k) & \gamma_{os}(k) \\
 {\gamma_{oe}}^*(k) & \omega_{even}(k) & \gamma_{es}(k)   \\
\gamma_{os}^*(k) & \gamma_{es}^*(k) & \omega_{substrate}(k)  \\
 \end{array} } \right|
\label{MATRIX}
\end{equation}
Within the CMT formalism (which is a first order perturbation theory), the coupling between $\gamma_{eo}(k)$ is proportional to the spatial overlap between the odd and even modes, weighted by the  perturbation $\Delta\epsilon(\vec{r})$ of the dielectric function $\epsilon(\vec{r})$ associated to the PhC structure, i.e.:
\begin{equation}
 \gamma_{oe}(k)=\int\int {\vec{E}^*}_{odd}(\vec{r}) \cdot \Delta\mathbf{\epsilon}(\vec{r}) \cdot \vec{E}_{even}(\vec{r})~ d\vec{r}
\end{equation}
In general, an arbitrary perturbation of the dielectric function $\epsilon(\vec{r})$ would also introduce the other diagonal and off-diagonal terms into eq. \ref{MATRIX}.
\begin{eqnarray}
\gamma_{oo}(k)=\int\int {\vec{E}^*}_{odd}(\vec{r}) \cdot \Delta\epsilon(\vec{r}) \cdot \vec{E}_{odd}(\vec{r})~ d\vec{r}  \nonumber \\
\gamma_{ee}(k)=\int\int {\vec{E}^*}_{even}(\vec{r}) \cdot \Delta\epsilon(\vec{r}) \cdot \vec{E}_{even}(\vec{r})~ d\vec{r}  \nonumber \\
\gamma_{ss}(k)=\int\int {\vec{E}^*}_{substrate}(\vec{r}) \cdot \Delta\epsilon(\vec{r}) \cdot \vec{E}_{substrate}(\vec{r})~ d\vec{r} \nonumber \\
\gamma_{os}(k)=\int\int {\vec{E}^*}_{odd}(\vec{r}) \cdot \Delta\epsilon(\vec{r}) \cdot \vec{E}_{substrate}(\vec{r})~ d\vec{r} \nonumber \\
\gamma_{es}(k)=\int\int {\vec{E}^*}_{even}(\vec{r}) \cdot \Delta\epsilon(\vec{r}) \cdot \vec{E}_{substrate}(\vec{r})~ d\vec{r} \nonumber \\
\label{Self}
\end{eqnarray}
As a consequence, the perturbation would not only introduce even-odd mode coupling, but would also mix these modes with the \textit{substrate} mode ($\gamma_{es}$, $\gamma_{os}$); as well as with any other modes -which are not considered here-; and induce a frequency offset of the single dispersion profile through the diagonal terms. Hence the result of the perturbation is uncertain. If $\gamma_{es}=0$ and $\gamma_{os}=0$, the dispersion is then easier to control, at least from a qualitative point of view.
\newline
\indent Hereafter we give a procedure to generate a perturbation $\Delta\epsilon(\vec{r})$ such that $\gamma_{eo} (k)$ is non-zero, but also dominates also substantially over the other terms. Let us first look at the integrand $\vec{E}^*_{odd}(\vec{r}) \cdot \Delta\mathbf{\epsilon}(\vec{r}) \cdot \vec{E}_{even}(\vec{r})$. The integral can be non-zero only if this is an even function. Considering that the planar PhC waveguide belongs to the group of symmetry $C_{2v}$, that statement is equivalent to requiring the integrand to have at least the element of symmetry $A_1$ (namely, the fully symmetric function). Conversely, if that condition is not respected, then integrand will have both positive and negative parts which would mutually cancel out.

\begin{figure}[htbf]
\includegraphics[width=6cm]{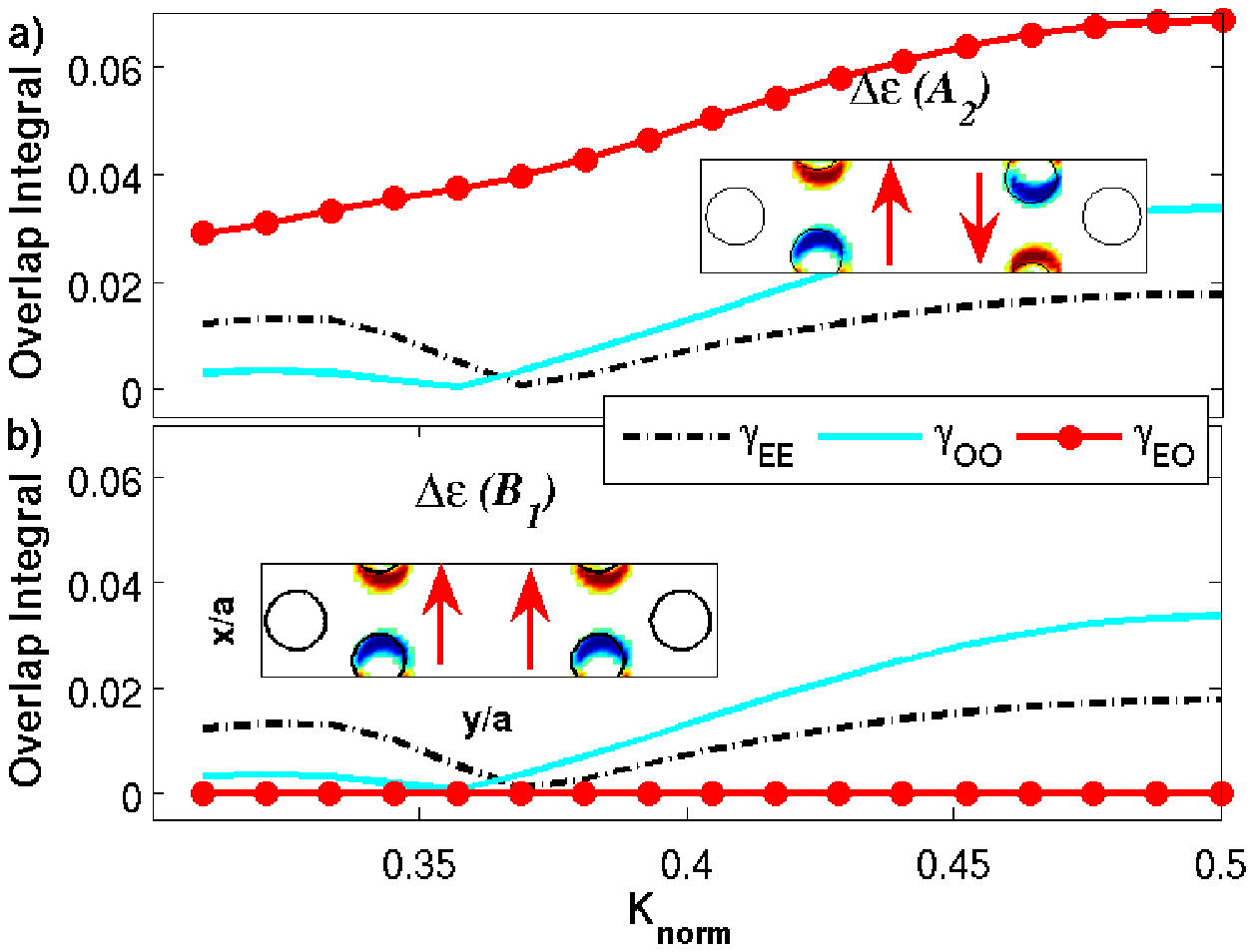} \includegraphics[width=6.5cm]{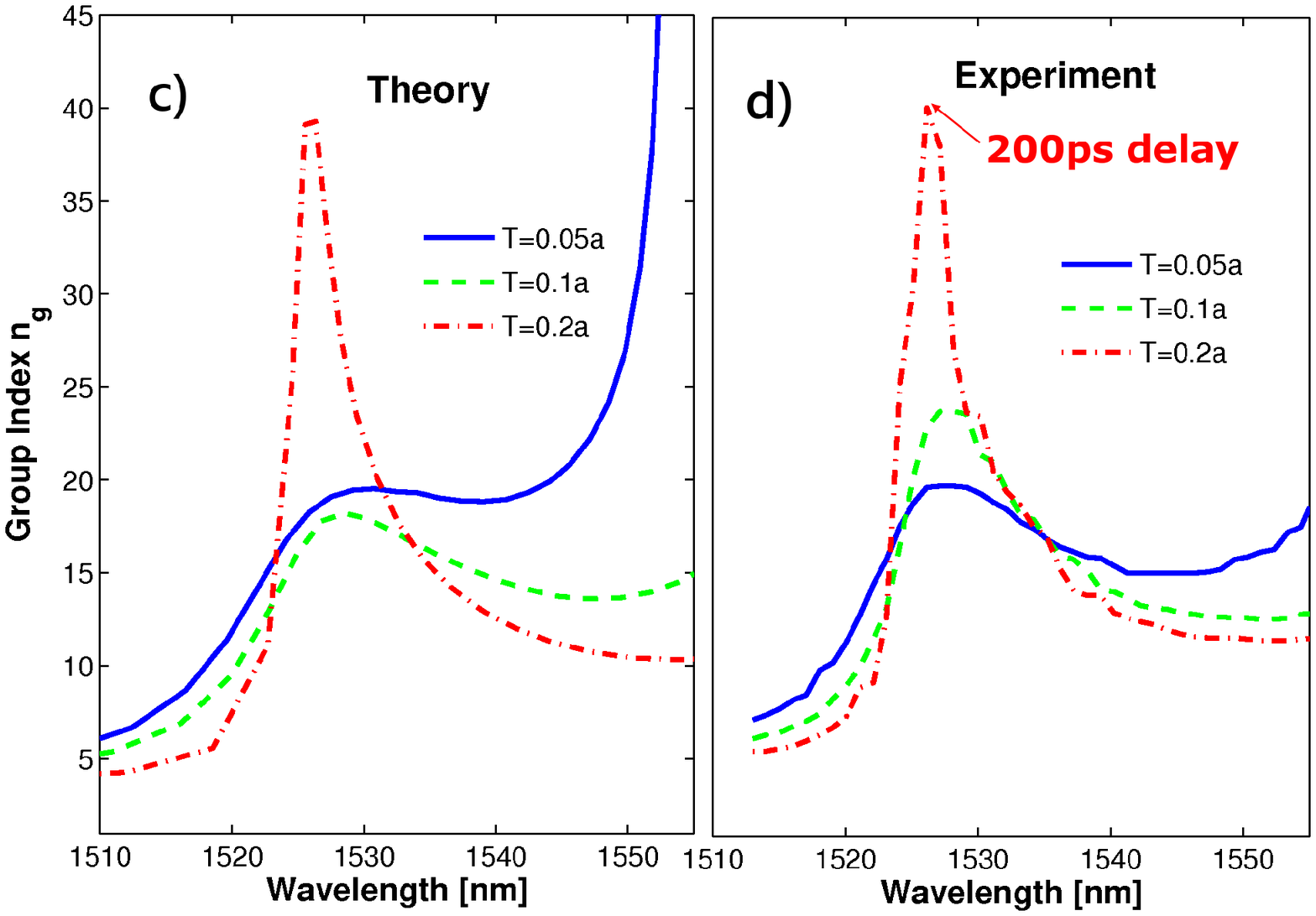}
\caption{a) Calculated $|\gamma_{ee}|$, $|\gamma_{oo}|$ and $|\gamma_{eo}|$ coefficients for T=0.15a. Inset : corresponding $\Delta\epsilon$ and its dominant symmetry (colors refers to opposite signs). b) Idem as a) but holes are moved in the same direction so the symmetry of $\Delta\epsilon$ is now $B_1$ instead of $A_2$. c) Group index dependence as a function of the wavelength calculated using MPB (resolution 18pts) for T= 0.05a, 0.1a, 0.2a. d) Corresponding measured dispersion in 1.5mm-long waveguides. The lattice period is $a=465nm$.}
\label{GammaCalcul} 
\end{figure}

\begin{table}
\centering
\caption{ C2v point group symmetry table } 
\begin{tabular}{|l||c|c|c|c|r|}
\hline
 &  E	&	$C_2$ (z)	&	$\sigma_v$ (xz)	&	$\sigma_v$ (yz)	& \\
\hline
$A_1$ &  1	&	1	&	1	&	1 & \\
$A_2$ &  1	&	1	&	-1	&	-1 & \\
$B_1$ & 1 &	-1	&	1	& 	  -1	&  \\
$B_2$	&	1	& 	-1	&	-1	&	1 & \\
\hline
\hline
$A_2 \otimes A_2$ & 1 & 1 & 1 & 1 & =$A_1$\\
$B_1 \otimes B_2$ & 1 & 1 & -1 & -1 & =$A_2$\\
\hline
\end{tabular}
\label{Tablec2v}
\end{table}
Proceeding similarly as in Ref. \cite{Painter:2003}, the Bloch modes in a PhC with a 2D hexagonal lattice can be labeled according to the table \ref{Tablec2v}. Based on the mode field distribution of the unperturbed structure in Fig. \ref{Dispersion}d, we conclude that the even and the odd modes belong to the symmetry subgroups $B_1$ and  $B_2$ respectively. The symmetry condition on the integrand in $\gamma_{oe}(k)$ is then formulated as:
\begin{equation}
 A_1 = B_2 \otimes X \otimes B_1
  \label{groupeq}
\end{equation}
where $X$ is the symmetry of $\Delta\epsilon(\vec{r})$, still to be determined. It is found that $X$ must have the symmetry $A_2$. This implies that the dielectric perturbation is anti-symmetric with respect both to the $x=0$ and the $y=0$ symmetry planes. This also implies that
$\gamma_{oo}(k)=\gamma_{ee}(k)=\gamma_{ss}(k)=\gamma_{se}(k)=0$, i.e., the only non-zero elements are $\gamma_{eo}(k)$ and $\gamma_{so}(k)$. Indeed, since the even and the substrate modes have the same symmetry, they are not coupled together by a such perturbation. It can be observed that the coupling $\gamma_{so}(k)$ between the odd and the substrate modes is minimized if the initial frequency spacing between the two modes is large and if the changes of the dielectric function $\Delta\epsilon(\vec{r})$ is localized where the field of the even and odd modes is much stronger than that of the substrate mode. This is in general the case in the region close to the line defect (see fig. \ref{Dispersion}d).
\newline
\indent
Based on that analysis, a modified PhC geometry is generated by translating the first row of holes by $T/a=0.1$ along $x$ in opposite directions, as is shown in fig. \ref{Dispersion}b. It is straightforward to verify that the change introduced respects the $A_2$ symmetry (fig. \ref{GammaCalcul}a). As a consequence, the original symmetry of the waveguide is now broken. The corresponding dispersion (fig.\ref{Dispersion}c) now exhibits a kink near $\mathbf{K}=0.445$ and the lower (upper) band is translated to a lower (higher) frequency (spacing has now increased to $\delta f\,a/c=1.5\, 10^{-2}$). The origin of the kink (i.e. zero group velocity dispersion points) is related to the change from a crossing behaviour (even and odd uncoupled) to an anti-crossing one, which implies that the modes repel each other.
\par
 We like to stress that there are several ``odd'' and ``even'' modes in PhC (to avoid confusion the modes should rather be labeled according the $C_{2v}$ group) which have different symmetry. As a consequence the way we broke the symmetry depends on the initial design. Hence in a different design \cite{Hamachi:09}, the dispersion is controlled by coupling the even mode with the substrate mode (which is initially closer in frequency); and this is done using a perturbation with the symmetry $B_1$. In any case, the analysis based on group symmetry considerations gives access to the symmetry of the perturbation, and the location where it would be the most efficient. Therefore, we think this could help to reduce greatly the number of errors and trials done before eventually obtaining the desired dispersion.
\par
Because the waveguide symmetry is now broken, the strict classification of the modes as \textit{odd} or \textit{even} is not applicable. We therefore label the lower (upper) frequency mode as \textit{even-like} (\textit{odd-like}). The spatial distribution of the modes corresponding to these new bands is shown in fig. \ref{Dispersion}e and clearly reveals the lack of a well defined parity (odd or even). The modification of the spatial distribution of the modes, i.e. their hybridisation through mode coupling, can be connected to the dispersion.

In order to elucidate this point, we calculate the diagonal and off-diagonal terms introduced in (\ref{MATRIX}), which are shown in Fig. \ref{GammaCalcul}a. A closer look in the field distribution reveals that the perturbation still keeps some features with the $A_1$ symmetry and, as a result, small diagonal contributions $\gamma_{ee}$ and $\gamma_{oo}$ appear into the operator (\ref{MATRIX}). We repeated the same calculation with a perturbation with equal \textit{magnitude} but a different symmetry (namely $B_1$), as shown in Fig. \ref{GammaCalcul}b. Interestingly, while $\gamma_{ee}$ and $\gamma_{oo}$ are basically unchanged, the coupling between odd and even modes has disappeared. The dispersion (blue line on Fig. \ref{Dispersion}c ) has no zero GVD points or equivalently, no anti-crossing is observed. This comparison confirms the minor role played by the diagonal contributions which are only responsible for a moderate steepening of the dispersion around $\mathbf{K}=0.37$, but do not create any inflexion point.

It is now apparent that the features introduced in the dispersion are mainly related to the contribution $\gamma_{eo}$ above, which depends on $T/a$, a parameter defining the degree of asymmetry introduced in the structure. Therefore the change of the dispersion is controlled by an unique parameter.
\newline
\indent
Let us consider now the dispersion, namely the group index as a function of the wavelength, as the parameter $T$ is changed (fig. \ref{GammaCalcul}c). Indeed, while a flat-band over more then $10nm$ is obtained with T=0.05, the other values generate a dispersion where the group dispersion changes sign twice and a local maximum of the group index increasing sharply as T is further increased.


\indent We have fabricated 1.5mm-long PhC waveguide on a 190nm thick air-suspended GaInP membrane \cite{combrie_APL2009} with the above specified designwhere T is varied between 0.05a and 0.20a. The use of mode adapters reduces the coupling loss below 1.5dB/facet\cite{tran_APL2009}. The dispersion is measured using an interferometric technique\cite{parini:08}. This technique also allow investigating the effects due to the disorder (multiple scattering etc.); these effects were found negligible in our waveguides. The result is shown is Fig. \ref{GammaCalcul}d. The agreement with the predicted dispersion is very good. Residual differences (e.g. for $T/a=0.05$) can be attributed to the fabrication tolerances (e.g. the radius of the holes). The important point is that the transition from one of these dispersion profiles to another follows continuously the change in the parameter T; which can be further adjusted in order to reach the desired dispersion profile.
\newline 
\indent An important practical aspect in dispersion-engineered  waveguides is their robustness against disorder, particularly their ability in the regime of low group velocity. For instance, a group index of $40$ is reached for $T/a=0.2$ (fig. \ref{GammaCalcul}c-d). Thus a group delay of about 200ps was obtained on a $1.5$ mm long waveguide. This is quite a respectable target, comparable to the state of the art for delay lines based on nanophotonics\cite{vlasov2008,Shinya2008,melloni:10,Schultz:10}. Furthermore, in this slow light regime, the related increase of propagation losses is limited to about 5dB. For example, this allowed demonstrating efficient Four Wave Mixing in PhCs\cite{Colman:11_Ol}.
\newline \par
In conclusion, we have revisited the concept of dispersion tailoring based on mode coupling and we have implemented it in a single-mode line-defect PhC waveguide. The group theory analysis was used to determine a suitable perturbation of the PhC structure such that only the even and the odd modes are coupled. The change in the dispersion, and, particularly, the onset of two zero group dispersion points, is explained through the anti-crossing related to the even-odd mode coupling. A variety of dispersion profiles, including flat-band, is controlled through a single geometrical parameter, representing the degree of asymmetry which is introduced into the PhC structure. The same concept can also be extended to the coupling between two arbitrary modes (not only odd and even ones). This result is relevant to a broad range of applications, particularly parametric effects (e.g. Four-Wave Mixing) and it enables the optimization of the phase matching. We believe  that it could be applied to other systems as well (e.g. PCF) and enable an even more accurate control of the dispersion.
\bigskip
\par\indent This work has been supported by the 7th framework of the European commission through the GOSPEL project (www.gospel-project.eu). We thank M. Santagiustina, G. Eisenstein and S. Trillo for enlightening discussions.
\end{document}